\documentclass[12pt]{iopart}

\usepackage{iopams}

\def\to{\rightarrow}

\def\AJ{{\it Ap. J.} }

\def\ANJ{{\it Astron. J.} }

\def\CQG{{\it Class. Quantum Gravity} }

\def\GRG{{\it Gen. Relativity and Gravitation} }

\def\IJMP{{\it Int. J. Mod. Phys.} }

\def\MPL{{\it Mod. Phys. Lett.} }

\def\PL{{\it Phys. Lett.} }
\def\PR{{\it Phys. Rev.} }

 \def\be{\beta}  
   
   \def\ka{\kappa}
\def\la{\lambda}

  \def\mn{{\mu\nu}} \def\cl{{\cal L}}

 \def\frac#1#2{{\textstyle{{#1}\over
			{#2}}}} 
\def\lsim{\mathrel{\rlap{\lower4pt\hbox{\hskip1pt$\sim$}}
		\raise1pt\hbox{$<$}}}
\def\gsim{\mathrel{\rlap{\lower4pt\hbox{\hskip1pt$\sim$}}
		\raise1pt\hbox{$>$}}} \def\sqr#1#2{{\vcenter{\vbox{\hrule height.#2pt
				\hbox{\vrule width.#2pt height#1pt \kern#1pt \vrule width.#2pt} \hrule
				height.#2pt}}}}
\def\square{\mathchoice\sqr66\sqr66\sqr{2.1}3\sqr{1.5}3}
\def\beq{\begin{equation}} \def\eeq{\end{equation}}
\def\beqa{\begin{eqnarray}} \def\eeqa{\end{eqnarray}}

\usepackage{lineno,hyperref}
\usepackage{latexsym}
\usepackage{amsmath}
\usepackage{amssymb}
\usepackage{iopams}

\modulolinenumbers[5]

\begin{document}

\title[Visible and Dark Matter Profiles in a Non-Minimally Coupled Model]{Visible and Dark Matter Profiles in a Non-Minimally Coupled Model}

\author{An\'ibal Silva and Jorge P\'aramos}

\address{Departamento de F\'isica e Astronomia and Centro de F\'isica do Porto,\\Faculdade de Ci\^encias da Universidade do Porto, \\Rua do Campo Alegre 687, 4169-007 Porto, Portugal}
\ead{up201008538@fc.up.pt, jorge.paramos@fc.up.pt}
\vspace{10pt}
\begin{indented}
\item[]
\end{indented}

\begin{abstract}

A previous work found that a nonminimally coupled theory of gravity can, under appropriate conditions, give rise to an additional contribution to the field equations interpreted as dark matter \cite{DMOBJP}: in particular, the density of this dark matter component was found to scale as a power of the density of visible matter. However, no explicit solution for the modified field equations was provided, so that a direct computation of the specific density profile followed by visible matter is missing.
	
This question is now addressed: analytical solutions to the modified field equations are derived in the appropriate perturbative regime and characterised, with an emphasis on directly obtaining the visible matter profile in a self-consistent way. We compare with known profiles for visible and dark matter and obtain constraints on the parameters of the model.
	
\end{abstract}

% Uncomment for PACS numbers
%\pacs{00.00, 20.00, 42.10}
%
% Uncomment for keywords
\vspace{2pc}
\noindent{\it Keywords}: Modified gravity, Non-Minimal Coupling, Dark Matter
%
% Uncomment for Submitted to journal title message

\submitto{\CQG}
%
% Uncomment if a separate title page is required
\maketitle
% 
% For two-column output uncomment the next line and choose [10pt] rather than [12pt] in the \documentclass declaration
%\ioptwocol
%

\section{Introduction}
\label{sec:int1}

Einstein's field equations have been exhaustively studied throughout the past century, providing new ways of understanding our Universe, from the local to the cosmological scale (see {\it e.g.} Refs. \cite{Will2006,OBJP2012}). Notwithstanding, General Relativity (GR) does not fully account for observations, thus giving rise to the need to invoke exotic forms of dark energy \cite{DE} or dark matter \cite{DM}: indeed, one of the main problems regarding astrophysics is the riddle posited by the latter, which is inferred from the mismatch between gravitational profiles (inferred {\it e.g.} from rotation curves of galaxies) and an insufficient visible matter content.

Traditionally, these dark components are modelled by  additional fields with suitable properties; alternatively, one can resort to modifications of GR itself, such as the class of the well known $ f(R) $ theories, which generalize the Einstein-Hilbert action functional by replacing the linear dependence on the scalar curvature action \cite{felice}.

More generally, one can also insert a nonminimal coupling between an arbitrary function of the scalar curvature and the Lagrangian density of matter \cite{early,Bertolami:2007gv}: this can introduced phenomenologically or arise from one-loop vacuum-polarization effects in the formulation of Quantum Electrodynamics in a curved space-time \cite{PhysRevD.22.343} or from considering a Riemann-Cartan geometry \cite{Goenner:1984zx}. In recent years, this theory has been carefully studied and has yielded several interesting results \cite{DE1,DE2,DE3,Bertolami:2011fz,Bertolami:2013raa,Paramos:2012mm,CastelBranco201425,0264-9381-25-24-245017,PhysRevD.86.044034}, while avoiding potential pitfalls \cite{PhysRevD.79.104010,Bertolami:2014hra}  (see Ref. \cite{NMCreview} for a discussion).

The purpose of this work is to further develop a mechanism to mimic dark matter distributions in galaxies by resorting to a NMC model in a relaxed regime, as first described in Ref. \cite{DMOBJP}: this work found that, for a power-law NMC, $f_2(R) \sim R^n$, the dark matter density $\rho_{dm}$ scales as a power of the visible matter density $\rho$, $\rho_{dm} \sim \rho^{1/(1-n)}$.

Crucially, the specific density profile of the visible matter density $\rho = \rho(r)$ was not computed --- instead the aim was to correlate known visible matter profiles with those of dark matter, and infer the appropriate value for the exponent $n$ and the characteristic curvature scale of the NMC.

This work extends and completes Ref. \cite{DMOBJP} by solving the modified field equations in the assumed regime: this allows us to not only confirm the previously obtained scaling between visible and dark matter, but also to explicitly solve for all quantities --- namely $\rho(r)$, $\rho_{dm}(r)$ and the equation of state (EOS) parameter $\omega(r)$.

In what follows, we first give a general description of the dynamics of the model. Using suitable choices for the functions $ f_1(R) $ and $f_2(R)$ and taking into the account the dominance of dark matter in the outer regions of a galaxy, we then simplify the ensuing system of differential equations and provide an approximate analytical solution. This allows us to assess the impact of this solution in both geometrical and physics quantities. Finally, we compute the relevant observables by using data of rotational curves of galaxies, in order to constraint the parameters of the model.

\section{The model} \label{model}

The action of the NMC model under scrutiny is written as
\beq S=\int d^4x \sqrt{-g}\left[\ka f_1(R)+f_2(R)\cl\right]~~, \label{action}\eeq
where $ \kappa = c^4 / 16 \pi G$ and $\cl $ is the Lagrangian density of matter; notice that GR can be recovered by setting $ f_1(R) = R $ and $ f_2(R) = 1 $.

Variation with respect to the metric $ g_{\mu \nu} $ yields the modified field equations
\begin{eqnarray}
&& 2\left(\kappa F_1+F_2\cl\right)G_{\mu\nu} - \left[\kappa \left(f_1-F_1R\right)-F_2 R\cl \right] g_{\mu\nu}= \nonumber\\ &&f_2T_{\mu\nu}+2\Delta_{\mu\nu}\left(\kappa F_1+F_2\cl\right) ~~, \label{fieldequation}
\end{eqnarray}
where $ F_i \equiv df_i/dR$, $ (i=1,2)$, $ \Delta_{\mu\nu} \equiv \Box_{\mu \nu}- g_{\mu \nu} \Box $ is defined for convenience and $ T_{\mu \nu} $ is the energy-momentum tensor of matter, given by
\beq T_{\mu \nu} = -{2 \over \sqrt{-g}} {\delta(\sqrt{-g}\cl)\over \delta g^{\mu \nu}} ~~.\eeq
One of the most striking features of this model is that it is no longer covariantly conserved, since
\beq \label{noncons}\nabla^\mu T_\mn = {F_2 \over f_2}(g_\mn \cl - T_\mn) \nabla^\mu R~~, \eeq
may be non-vanishing (see Ref. \cite{noncons} for a thorough discussion).

In what follows, we shall consider that matter is described as an isotropic perfect fluid,
\beq T_{\mu\nu}=\left(\rho+p\right)u_\mu u_\nu + pg_{\mu\nu} \label{et0}~~, \eeq
where $\rho$ is the energy density, $p$ is pressure, and $u_\mu$ is the 4-velocity, which obeys $u_\mu u^{\mu}=-1$ and $ u_\mu u_{;\nu}^{\mu}=0 $.

\subsection{Relaxed Regime}

Following Refs. \cite{DMOBJP,PhysRevD.89.044012,unimodularNMC}, one now considers the additional constraint
\beq \kappa F_1+F_2\cl= \kappa \label{re0}~~, \eeq
which fixes the so-called relaxed regime for the NMC. We assume that this condition is valid only in the exterior region where dark matter dominates, but is not applicable to the inner galactic region. 

Through a suitable definition of variables, the restriction above can be seen as equivalent to a fixed point condition, since the $ \Delta_{\mu\nu} $ term on the field equations \eqref{fieldequation} vanishes. This appraisal is clarified through the equivalence discussed below.

\subsubsection{Equivalence with Two Scalar Field Theory}

Interestingly, condition (\ref{re0}) is straightforwardly interpreted in the light of the equivalence between the action (\ref{action}) of the model here considered and a two-scalar field theory \cite{0264-9381-25-24-245017} written, in the Einstein frame ({\it i.e.} where the scalar curvature is uncoupled), as
\begin{equation} \label{NMCequivaction} S = \int d^4x \sqrt{-g} \left[ \ka \left( R + {f_1(\phi) \over \psi^2} - { \phi \over \psi} - {3 \over 2\psi^2} \nabla_\mu \psi \nabla^\mu \psi \right) + f_2(\phi) \cl\left({g^\mn \over \psi},\chi\right) \right] ~~.\end{equation}
Notice that the matter Lagrangian is coupled to both scalar fields $\phi$ and $\psi$: the former through $f_2(\phi)$ and the latter through the physical metric $g^\mn / \psi$ used to construct $\cl$ ({\it e.g.} when building kinetic terms for the matter fields $\chi$).

Variation of the action with respect to the two scalar fields yields the dynamical identification
\begin{equation} \phi = R~~~~,~~~~\psi = F_1(R) + {1 \over \ka} F_2(R) \cl ~~,\end{equation}
so that, while $\phi$ acts as an auxiliary field with no kinetic term, $\psi$ embodies an additional scalar degree of freedom, as found in $f(R)$ theories \cite{scalarfR}.

Thus, condition (\ref{re0}) corresponds to solutions of the form $\psi = 1 $ and can be interpreted as an asymptotic regime for this dynamical scalar field, providing a physical rationale for this assumption.

In a cosmological context, it was found that condition \eqref{re0} indeed arises naturally from a dynamical system formulation \cite{dynamical,PhysRevD.94.064036}, allowing for a de Sitter expansion of the Universe in the presence of a non-negligible matter content. Furthermore, it can lead to interesting phenomenological modifications of the Friedmann equation \cite{PhysRevD.89.044012} (see Ref. \cite{DE1} for similar results).

More recently \cite{unimodularNMC}, it was shown that the condition above gives rise to a model which closely resembles unimodular gravity \cite{unimodular}, without the eponymous need for {\it a priori} fixing of the determinant $\sqrt{-g} = 1$: inserting \eqref{re0} into the field equations \eqref{fieldequation} leads to the simplified form
\beq R_{\mu\nu}= {1 \over 2\kappa}(f_2T_{\mu\nu}+\kappa f_1 g_{\mu\nu})~~,\label{fe0}\eeq

with trace
\beq R={1 \over 2\kappa}\left(f_2T+4\kappa f_1 \right)~~. \label{tr0} \eeq

Solving for $ f_1 $ and replacing into \eqref{fe0} leads to the traceless form

\beq R_{\mu\nu}-{1 \over 4}Rg_{\mu\nu}={1 \over 2\kappa }\left(T_{\mu\nu}-{1 \over 4}Tg_{\mu\nu}  \right)f_2~~, \label{fe3} \eeq
which is strikingly similar to Unimodular gravity, although complemented by the NMC on the r.h.s..

Similarly to the constraint $\sqrt{-g}=1$ found in Unimodular gravity, condition (\ref{re0}) could be enforced in the action (\ref{action}) by including include a suitable Lagrange multiplier: the natural step, however, would be to further promote this Lagrange multiplier to an additional degree of freedom with an appropriate kinetic and potential term --- which, following the preceding discussion, would be equivalent to a rather convoluted theory with three scalar fields.

Furthermore, we only assume the validity of Eq. (\ref{re0}) in the outer region of galaxies, where dark matter dominates: this is incompatible with a ``hardcoded'' constraint at the action level, but follows naturally from the proposed interpretation --- that the former is not fundamental, but dynamically attained from random initial conditions for the matter distribution. Notwithstanding, the similarity between Unimodular gravity and an $f(R)$ theory with a NMC in the relaxed regime is suggestive and warrants further studies.

In an astrophysical context, condition \eqref{re0} was used as a way to mimic dark matter, in both clusters  \cite{PhysRevD.86.044034} and galaxies \cite{DMOBJP}. In the latter, condition \eqref{re0} provides an immediate relation for $\rho = \rho(R)$ in the case of a pressureless dust distribution; that study also showed that numerical solutions to the overall field equations admit solutions that oscillate around the form specified by Eq. \eqref{re0}.

In general, Eq. \eqref{re0} provides an additional constraint $R = R(\cl) $ to the ensuing system of differential equations of motion: as such, an extra EOS is no longer required {\it a priori} to close it, as one may in principle solve it completely --- and then read the corresponding (non-constant) EOS parameter $\omega(r) = p(r)/\rho(r)$.

\subsection{Conformal Transformation} \label{scaling}

Notice that one could always insert a dimensionless constant $ \Omega $ in the r.h.s. of the relaxed condition \eqref{re0} and conformally transform it away: to see this clearly, we write it in terms of a scaled curvature $ \tilde{R} $,

\beq \kappa {df_1 \over d\tilde{R} }+{df_2 \over d\tilde{R}} \cl = \kappa \Omega \to  {df_1 \over d(\Omega \tilde{R}) }+{df_2 \over d(\Omega \tilde{R})} \cl = \kappa~~,  \label{re1} \eeq

\noindent so that making the identification $R = \Omega \tilde{R} $ leads us back to condition \eqref{re0}: this is attained by conformally scaling the metric through $\tilde{g}_{\mu\nu} = \Omega g_{\mu \nu}$.  We denote $ g_{\mu \nu} $ as the the conformal metric, which will be used to solve our field equations, and then transform back to the physical metric $ \tilde{g}_{\mu \nu} $.

Following Refs. \cite{Brown:1992kc,perfectfluids}, where it was argued that the Lagrangian density of a perfect fluid takes the form $\cl = -\rho $, Ref. \cite{DMOBJP} adopts trivial $f_1(R)$ and power-law $f_2(R)$ forms
\beqa f_1(\tilde{R}) &=& \tilde{R}~~, \nonumber \\ f_2 (\tilde{R}) &=& 1 + \left( {\tilde{R} \over R_n}\right)^n~~, \label{fchoice} \eeqa
where $ R_n $ is a characteristic curvature scale and $ n $ is a power-law exponent: in order to mimic dark matter profiles, the latter must be negative, as we expect it to dominate in the outer regions of the galaxy when the curvature becomes sufficiently small. 

One should notice that the above choice has also been used in a cosmological setting \cite{DE1}, based upon a similar reasoning: to drive dark energy, a negative power-law NMC should be used: while negligible in the early Universe, it becomes dominant as the scalar curvature drops at late times. However, the reported numerical fits to existing cosmographic studies for the evolution of the deceleration parameter \cite{Gong} lead to large negative exponents, $n\sim -10$, incompatible with the visible and mimicked dark matter profiles considered in this study.

To account for this, one usually assumes that both functions $f_i(R)$ can be in general written as a Laurent series
\begin{equation} f^i(R) = \sum_{j=-\infty}^{+\infty} \left( {R \over R_{ij}}\right)^j ~~.\end{equation}
Thus, the adoption of a simpler power-law form for $f_i(R)$ in a particular context assumes that, for the values of the scalar curvature relevant in that scenario, one of the terms of the Laurent series above is dominant, $f_i(R) \sim (R/R_{in})^n$ (this argument was previously invoked in $f(R)$ theories \cite{felice}).

Replacing the functions (\ref{fchoice}) in the (rescaled) relaxed regime condition \eqref{re1} yields 
\beq \ka {df_1 \over d\tilde{R}} + {df_2 \over d\tilde{R}} = \ka - {n \over R_n} \left({\tilde{R} \over R_n}\right)^{n-1} \rho = \Omega \ka~~, 
\eeq
and solving for $ \tilde{R} $ enables the relation
\beq \label{new} \tilde{R} = R_n \left({n \over \kappa(1-\Omega)}{ \rho \over R_n} \right)^{1/(1-n)}~~.    \eeq

Ref. \cite{DMOBJP} obtained a similar result, albeit in a less formal fashion: it adopted the same form for the relevant functions (\ref{ffunctions}) and considered visible matter with vanishing pressure $\omega =0$, so that the trace of the modified equations of motion (\ref{fieldequation}) reads
\beq \label{traceeq} \tilde{R} = (1-2n)\left( {\tilde{R} \over R_n}\right)^n {\rho \over 2\ka} - {3n \over \ka} \square\left[\left( {\tilde{R} \over R_n}\right)^{n-1}{\rho  \over R_n}\right] ~~,\eeq
assuming that the mimicked dark matter component dominates and the contribution to the r.h.s. from GR may be neglected.

Inspection then shows that, if one assumes that the cumbersome derivative term vanishes,
\beq \left( {\tilde{R} \over R_n}\right)^{n-1}{\rho  \over R_n} = \rm{const}. ~~, \eeq
which is equivalent to the constraint (\ref{re0}), the trace equation has the self-consistent solution
\beq \tilde{R} = R_n\left({1-2n\over 2 \kappa}{\rho \over R_n}\right)^{1/(1-n)}~~. \label{curvaturetilde} \eeq
A subsequent numerical analysis showed that the general, unconstrained solutions to Eq. (\ref{traceeq}) do not obey the constraint (\ref{re0}), but exhibit negligible oscillations around this solution --- validating the aforementioned interpretation of condition (\ref{re0}) as a relaxed regime attained dynamically.

Comparing Eqs. (\ref{new}) and (\ref{curvaturetilde}) and matching these two results shows that the conformal factor relating the two metrics is
\beq \Omega =  {1-4n \over 1-2n}~~. \eeq
As expected, GR is recovered by setting $n=0$, implying that $ \Omega=1 $.

\subsection{Dark Matter Mimicking} \label{DMenergytensor}

As stated, the main objective of this work is to mimic dark matter through a NMC between curvature and visible matter. Given the relation $ R = R(\cl) $ that arises out of the relaxed regime \eqref{re0}, we thus interpret the additional contributions to the field equations \eqref{fe3} as the energy density and pressure of an effective dark matter fluid, and write the field equations as the usual Einstein field equations with an additional energy-momentum tensor for dark matter,
\beq 2 \kappa G_{\mu \nu} = T_{\mu \nu} + T_{\mu \nu}^{(dm)}~~. \label{fe1} \eeq
Assuming that the latter also behaves as a perfect fluid,
\beq T_{\mu \nu}^{(dm)} = (\rho_{dm}+p_{dm}) v_\mu v_\nu + p_{dm} g_{\mu\nu}~~, \label{dmtensor} \eeq
with $ v_\mu = u_\mu $ (so that this mimicked dark matter is ``dragged'' by visible matter), one can easily arrive at
\begin{eqnarray}
p_{dm} &=& (f_2 -1)\omega \rho -{f_2 T \over 4} - {\kappa R \over 2}~~, \nonumber \\
\rho_{dm} &=& (f_2-1) \rho + {f_2 T \over 4} + {\kappa R \over 2}~~. \label{dmcomponents}
\end{eqnarray}

In the case of GR, these naturally vanish due to the usual trace condition $R = - T/2\ka$ and the minimal coupling $f_2(R) = 1$.

\section{Spherically symmetric, static case}\label{stationary}

In this section one aims to obtain a system of differential equations to our quantities of interest. Assuming spherical symmetry and stationarity, we adopt a Birkhoff metric of the form,
\beq ds^2 = -e^{2\phi(r)}dt^2 + e^{2\lambda(r)}dr^2 + r^2d\Sigma^2~~. \label{metric0} \eeq

The non-vanishing components of the energy-momentum tensor of matter \eqref{et0} can be easily evaluated,

\beq T_{tt}=\rho e^{2\phi} , ~~ T_{rr}=pe^{2\la}, ~~ T_{\theta \theta}=pr^2~~, \eeq

\noindent with trace $ T=3p-\rho$.

Given that a perfect fluid is described by a Lagrangian density $\cl = -\rho$ \cite{Brown:1992kc,perfectfluids}, while radiation obeys $\cl = p$, we adopt the general description

\beq \cl = - \gamma \rho =
\begin{cases}
	-\rho &, \gamma =1 \\
	p &, \gamma = -\omega
\end{cases} ~~.\eeq

Taking the radial component of Eq. \eqref{noncons}, we find
\beq \phi' (\rho + p)={F_2 \over f_2} (\cl-p)R'-p' ~~, \eeq
so that
\beq \phi' = -\beta{F_2 \over f_2}R'-{1\over \omega+1}\left(\omega' + \omega{\rho'\over\rho}\right)~~, \label{extf1}\nonumber
\eeq 
with the binary parameter 
\beq \beta = {\gamma + \omega \over 1+ \omega} = 
\begin{cases}
	1 &, \gamma =1 \\
	0 &, \gamma = -\omega
\end{cases} ~~,\eeq
introduced for convenience.

We now address the field equation \eqref{fe3}, resorting to the relation
\beq g^{tt}R_{tt}-g^{rr}R_{rr}=-{2 \over r}e^{-2\la}\left(\phi'+\la'\right)~~, \eeq
which, for the adopted diagonal metric, implies that
\beq e^{-2\la}\left({\phi'+\la' \over r}\right)={\omega+1\over4\ka}f_2\rho~~. \label{sum0} \eeq
The $ \theta-\theta $ component of \eqref{fe3} reads,
\beq {e^{-2\la}\over r}\left(\la'-\phi'\right)- {e^{-2\la} \over r^2} +  {1\over r^2} = {R \over 4} + {\omega + 1 \over 8\kappa} f_2\rho~~, \label{tt} \eeq
and solving Eq. \eqref{sum0} for $ \la'(r) $ yields
\beq \la' = {1+\omega \over 4\kappa }f_2\rho e^{2\la}r-\phi'~~. \label{dercomp1} \eeq

Inserting this result into Eq. \eqref{tt} leads to
\beq e^{2\la} = {1+2r\phi' \over 1- {r^2 \over 4}\left(R-{1+\omega \over 2 \kappa }f_2 \rho\right)}~~. \label{comp1} \eeq
Taking into the account the expression for the scalar curvature
\beq {e^{2\la} \over 2}R={e^{2\la} - 1 \over r^2}+\left({2 \over r} +\phi'\right)\left(\la'-\phi'\right)-\phi''~~, \label{scalcurv1} \eeq
and replacing the two preceding expression for $e^{2\la}$ and $ \la' $ produces
\beqa
&&(1+2r\phi')\left[{1+\omega \over 2 \kappa }f_2 \rho \left(1+ {r \over 2}\phi'\right)-{R\over 2}+{1 \over r^2}\right] = \\ &&  \left[{1 \over r^2}+{4 \over r} \phi' +2(\phi')^2+\phi''\right]  \left[ 1-  {r^2 \over 4}\left(R-{1+\omega \over 2 \kappa }f_2 \rho\right)\right]~~. \nonumber \eeqa
Simplifying both sides, one can write
\begin{eqnarray}\label{29}
{R\over 8} &-& {3 \over 8}{1+\omega \over \kappa}f_2 \rho r\left({1\over2r}+\phi'\right)+{ \phi' \over r} -  {1+\omega \over 8\kappa }f_2\rho r^2\left((\phi')^2-{\phi'' \over 2}\right) \\ \nonumber &+&  \left(1-{R \over 4}r^2\right)\left((\phi')^2+{\phi'' \over 2}\right) = 0~~.
\end{eqnarray}
Notice that, for consistency, one could obtain the same expression by differentiating Eq. \eqref{comp1} and then inserting Eq. \eqref{dercomp1}.

Differentiating $ \phi' $ yields
\begin{eqnarray} 
\phi '' &=& \beta\left[\left({F_2 \over f_2}\right)^2 (R')^2-{F_2' \over f_2}(R')^2-{F_2 \over f_2}R''\right] + {\omega' \over (1+\omega)^2}\left[\omega'+\omega{\rho' \over \rho}\right]\\ &-& {1 \over 1+ \omega}\left[\omega''+\omega ' {\rho' \over \rho} - \omega \left({\rho' \over \rho}\right)^2+\omega {\rho'' \over \rho}\right] \nonumber ~~,  
\end{eqnarray}
\noindent and replacing $ \phi' $ and $ \phi'' $ into Eq. \eqref{29}, we finally arrive at the cumbersome expression below,

\begin{eqnarray}
-{R \over 8}& + & {3 \over 8}{1+\omega \over \kappa}f_2 \rho r\left[{1\over 2r} - \beta{F_2 \over f_2}R'-{1 \over \omega+1}\left(\omega'+\omega{\rho' \over \rho}\right)\right] + {1\over r} {1 \over \omega + 1}\left(\omega'+\omega{\rho'\over \rho}\right) \nonumber  \\ &+& {\beta \over r}{F_2 \over f_2}R' + {1+\omega \over 8 \kappa} f_2 \rho r^2 \left[\beta\left(\beta - {1\over 2}\right)\left({F_2 \over f_2}\right)^2(R')^2  + {\beta \over 2}{F_2' \over f_2}(R')^2 + {\beta\over 2}{F_2 \over f_2}R''\nonumber \right.\\  &+& \left. {1\over 2}{\omega \over 1+\omega}{\rho'' \over \rho} + {1+4 \omega \over 2(1+\omega)^2}{\rho' \over \rho} \omega'+{(\omega')^2+(1+\omega)\omega'' \over 2(1+\omega)^2}+\omega {\omega - 1 \over 2(\omega+1)^2}\left({\rho' \over \rho}\right)^2 \nonumber \right. \\ &+& \left. 2\beta {1 \over 1+\omega} {F_2 \over f_2}R' \left(\omega{\rho' \over \rho} + \omega'\right)\right] - \left[1-{R \over 4}r^2\right]\left[\beta\left(\beta + {1\over2}\right)\left({F_2 \over f_2}\right)^2(R')^2 \nonumber \right. \\ &-& \left. {\beta \over 2}{F_2' \over f_2}(R')^2 - {\beta \over 2} {F_2 \over f_2}R'' + \omega {3\omega +1 \over 2(\omega+1)^2}\left({\rho' \over \rho}\right)^2 + {3(\omega')^2-(1+\omega)\omega''\over 2(1+\omega)^2}  \nonumber\right. \\ &-& \left.  {1\over 2}{\omega \over 1+\omega}{\rho '' \over \rho} + {4\omega-1 \over 2(1+\omega)^2}\omega' {\rho' \over \rho}+2\beta {1 \over 1+\omega} {F_2 \over f_2}R' \left(\omega{\rho' \over \rho}+\omega'\right)\right]=0~~, \label{asd}
\end{eqnarray}

\noindent a second order and non-linear differential equation for $ \rho(r)$, $\omega(r) $ and $ R(r) $.

\subsection{Metric components}
\label{metriccomponents}

Recall that in Section \ref{DMenergytensor} we wrote the modified equations of motion \eqref{fe0} as formally equivalent to the Einstein equations \eqref{fe1}, with visible matter supplemented by an effective dark matter component --- which actually depends on the scalar curvature $R$ and the Lagrangian density $\cl$ .

Notwithstanding, once the profiles $R(r)$ and $\cl(r)$ are known, this effective dark matter component is fully characterised and the metric elements are simply given by the standard interior solution of a spherical object with the energy momentum tensor \eqref{et0}:
\begin{eqnarray}\label{metrics}
m(r) &=& 4 \pi \int \tilde{\rho}_t(r) r ^2 dr~~,~~ e^{2 \la(r) } = \left(1-{m(r) \over 8 \pi \kappa r}\right)^{-1}~~, \nonumber \\
e^{2 \phi(r)}  &=& \exp\left(\int {4 \pi \tilde{p}_t(r) r^3 + m(r) \over r(8 \pi \kappa r - m(r))} dr\right)~~,
\end{eqnarray} 
\noindent where $ \tilde{\rho}_t $ and $ \tilde{p}_t $ denote, respectively, the total (visible + dark) energy density and pressure in the physical frame. Notice that one could always compute the metric elements by using the expressions found in Section \ref{stationary}: however, the above is clearly more convenient, once the mimicked dark matter profile has been written explicitly. Furthermore, it is known that a galaxy is within the validity of the Newtonian regime, $g^\mn \sim \eta^\mn$.

\section{Analytical Solution and Application} \label{analytical}

The above differential equation \eqref{asd} may be simplified by suitably defining the dimensionless quantities
\beq x \equiv r \sqrt{R_2} ~~,~~y \equiv {R \over R_2}~~,~~\varrho \equiv {\rho \over 2\ka R_2} ~~,\eeq
\noindent where $R_2$ is an arbitrary scale. Given this, the relaxed regime condition \eqref{re0} and trace Eq. \eqref{tr0} can be written as
\begin{eqnarray} \label{reducedconditions} 
1 &=& F_1(y)- 2\gamma F_2(y) R_2\varrho~~, \\ y &=& f_2(y)(3\omega-1)\varrho + {2 f_1(y)\over R_2} ~~, \nonumber  \end{eqnarray}
\noindent while the differential equation \eqref{asd} yields:

\begin{eqnarray}
& & 3 (1+\omega) f_2 \varrho x\left[{1\over x} - 2\beta{F_2 R_2 \over f_2}y'- {2 \over \omega+1}\left(\omega'+\omega{\varrho'\over\varrho}\right)\right] + {8 \be \over x }{F_2 R_2 \over f_2}y' \nonumber \\  &+& {8 \over x}{1 \over \omega+1}\left(\omega'+\omega{\varrho'\over\varrho}\right) + (1+\omega)f_2 \varrho x^2 \left[\beta\left(2\beta - 1\right)\left({F_2 R_2\over f_2}\right)^2(y')^2 \nonumber \right. \\ &+& \left. \beta {F_2' R_2^2 \over f_2}(y')^2 + \beta {F_2 R_2\over f_2}y'' + {(\omega')^2+(1+\omega)\omega'' \over (1+\omega)^2} +  \omega {\omega -1 \over (\omega+1)^2}\left({\varrho' \over \varrho}\right)^2  \nonumber \right. \\ &+& \left.  {\omega \over 1+\omega}{\varrho'' \over \varrho} + {1+4 \omega \over (1+\omega)^2}{\varrho' \over \varrho} \omega' + 4\beta {1 \over 1+\omega} {F_2 R_2\over f_2}y' \left(\omega{\varrho' \over \varrho}+\omega'\right)\right] \nonumber \\ &-& \left[4 -y x^2\right]\left[\beta\left(2\beta + 1\right)\left({F_2 R_2\over f_2}\right)^2(y')^2 - \beta {F_2' R_2^2 \over f_2}(y')^2 -  \beta {F_2 R_2\over f_2}y'' \nonumber \right. \\ &+& \left. {3(\omega')^2-(1+\omega)\omega''\over (1+\omega)^2} + \omega {3\omega +1 \over (\omega+1)^2}\left({\varrho' \over \varrho}\right)^2 - {\omega \over 1+\omega}{\varrho '' \over \varrho}+ {4\omega-1 \over (1+\omega)^2}\omega' {\varrho' \over \varrho}\nonumber \right. \\ &+& \left. 4\beta {1 \over 1+\omega} {F_2 R_2\over f_2}y' \left(\omega{\varrho' \over \varrho}+\omega'\right)\right]-y=0~~.\label{asd2}
\end{eqnarray} 

Recalling the discussion of paragraph \ref{scaling} and following Ref.\cite{DMOBJP}, we adopt the ``conformally transformed'' forms
\beq \label{ffunctions}
f_1(R) = {1-2n \over 1-4n} R,~~~~    
f_2(R) = 1 + \left( {1-2n \over 1-4n}{R \over R_n}\right)^n  ~~, 
\eeq
so that conditions \eqref{reducedconditions} read
\begin{eqnarray} \label{reducedconditions1} 
y^{1-n} &=& (1-4n) \varrho ~~, \\ \nonumber y &=& (1-4n) ( 1 + y^n) (1-3\omega)\varrho~~,  \end{eqnarray}
having imposed the natural identification $R_2 = R_n(1-4n)/(1-2n)$. This allows us to write the set of relations 
\begin{eqnarray}\label{f2andy} f_2 &=& 1 + y^n~~, \nonumber\\ {F_2 R_2 \over f_2} &=& n {y^{n-1} \over 1+ y^n}~~, \\ {F_2' R_2^2 \over f_2} &=& (n-1)n {y^{n-2} \over 1 + y^n} \nonumber ~~,\end{eqnarray}
so that the differential equation \eqref{asd2} only involves the exponent $ n $.

Solving the trace in Eq. \eqref{reducedconditions1} yields
\beq y^n = [(1-4n)\varrho]^{n/(1-n)} = {1\over 3\omega} -1 ~~. \label{reducedomega} \eeq
Since visible matter is known to be non-relativistic, thus behaving as a perfect fluid with negligible pressure ({\it i.e.} dust), we aim for a very small EOS parameter, $\omega \sim 0$. From the above, this implies a strong NMC, $ f_2 \approx y^n \gg 1 $; furthermore, as we consider a negative exponent $n<0$, this condition is equivalent to a very small reduced curvature $y \ll 1$ --- prompting the perturbative treatment to follow.

Our visible quantities are written implicitly in terms of scalar curvature as:
\beq \rho = {1 \over 1-4n}2 \kappa R_2 \left(R \over R_2\right)^{1-n}, ~~ \omega= {1 \over 3f_2}= {1 \over 3\left[1+\left(R \over R_2\right)^n\right] }\approx {1\over 3}\left(R \over R_2\right)^{-n} \label{rhoandomega}~~, \eeq
while the mimicked dark matter distribution can be read by replacing the conformally transformed functions \eqref{ffunctions} in Eq. \eqref{dmcomponents}:

\beq \label{dmvariables1}
\rho_{dm} ={1-n \over 1-4n}2 \kappa R,~~~~~~  p_{dm} = {n  \over 1-4n}2 \kappa R\left[1+{1 \over 3n}\left(R \over R_2\right)^{-n}\right]~~.  
\eeq

Asides from the conformal factor $ \Omega $, these expressions matches the one previously found in Ref. \cite{DMOBJP} --- with an additional perturbative contribution to $p_{dm}$ due to the pressure of visible matter (which in that work was taken to vanish exactly, $w =0$).

We now attempt to solve the differential equation \eqref{asd2} by resorting to the reduced expressions for visible matter \eqref{reducedconditions1}, EOS parameter \eqref{reducedomega} and the expressions which relate the coupling function $ f_2 $ and the reduced curvature $ y $ \eqref{f2andy}. To do so, we first solve it for $ y'' $ and, since $y \ll 1$, expand \eqref{asd2} perturbatively to first order, yielding the considerably simplified equation,
\beq y''(x) + {2\over x}y'(x) - (1+2n) {y'(x)^2 \over y(x)}  = 0 \label{difeq}~~. \eeq

This non-linear second order differential equation has an analytical solution given by
\beq y(x) = y_f \left({ 2x \over x_f + x}\right)^{1\over 2n} \label{redcurv}~~. \eeq
where $x_f$ and $y_f$ are constants: it is easy to check that at $ x = x_f $, $ y(x_f) =y_f $, so this normalization simply gives us the value of the scalar curvature evaluated at a given radius $ x_f $: we choose to identify $ x_f $ as the reduced radius of a galaxy, while $ y_f $ is the corresponding value of the reduced scalar curvature.

In the asymptotic regime $ x \gg x_f $, {\it i.e.} for a region sufficiently far away from the galaxy, one has
\beq y(x) \sim  2^{1\over 2n} y_f\left(1- {1\over 2n} {x_f\over x}\right)~~,\label{aprox2} \eeq
so that the scalar curvature approaches the value
\beq y(\infty) \sim 2^{1\over 2n} y_f~~, \eeq

\noindent which we interpret as the background contribution of the curvature.

Using solution \eqref{redcurv}, the scalar curvature and the respective dark matter contributions can be explicitly written as
\begin{eqnarray} \label{explicit}
R(r) &=& R_f \left(2r \over r_f + r\right)^{1\over2n}~~, \\
\rho_{dm} (r) &=& {1-n \over 1-4n}2 \kappa R_f\left( 2r \over r_f + r\right)^{1\over2n}~~, \nonumber \\ 
p_{dm}(r) &=& {n \over 1-4n} 2 \kappa R_f\left(2r \over r_f + r\right)^{1\over2n}  \left(1+ {1\over 3\sqrt{2}n}\left(R_f \over R_2 \right)^{-n}\sqrt{1 + {r_f \over r}}\right) ~~ \nonumber,
\end{eqnarray}
while the corresponding visible matter quantities read
\beq
\rho (r) = {2 \kappa R_f \over 1-4n}\left(R_f \over R_2\right)^{-n}\left(2r \over r_f+r\right)^{1-n \over 2n},~~  
\omega(r) = {1\over 3\sqrt{2}}\left(R_f \over R_2 \right)^{-n}\sqrt{1 + {r_f \over r}}~~,
\eeq
where $ r_f $ denotes the radius of a given galaxy and $ R_f $ the corresponding scalar curvature.

\subsection{Energy conditions}

Before proceeding, we now assess if the obtained result do not break the desired criteria for physical solution, namely the energy conditions and the absence of the so-called Dolgov-Kawasaki instabilities. Following Refs. \cite{DMOBJP,PhysRevD.79.104010}, the strong, null, weak, and dominant energy conditions (SEC, NEC, WEC and DEC, respectively) are obeyed if 

\beq 1 + { E_n \over 1 + \left( {R \over R_0} \right)^{-n} } \geq 0 ,\eeq
with 
\beq E_n = \begin{cases}-2n &, ~~{\rm SEC} \\ 0 & , ~~{\rm NEC} \\ 2n & , ~~{\rm DEC} \\ n & , ~~{\rm WEC} \end{cases}~~.\eeq

Since the NMC is strong, $f_2(R) \sim (R/R_n)^n \gg 1$, the above is satisfied if $1 + E_n > 0$. Thus, the NEC is trivially satisfied and, since the exponent $n$ is negative, so is the SEC. The DEC and WEC are obeyed if $ -1/2< n < 0$, which is the case for all the relevant visible or dark matter profiles considered here.

We now ascertain whether the considered model does not give rise to the so-called Dolgov-Kawasaki instabilities, corresponding to the undesired exponential growth of initially small curvature perturbations \cite{Dolgov:2003px}. Following Ref. \cite{Faraoni:2007sn} (see also Refs. \cite{PhysRevD.79.104010,DKNMC}), the former is avoided if the associated mass scale $m_{DK}$ is real-valued,
\beq m_{DK}^2 = { \kappa (F_1 + F_1' R) + (F_2 + F_2' R ) \cl + F_2 T\over 3 (\kappa F_1' + F_2' \cl) } >0 ~~.\eeq
Given the adopted forms (\ref{ffunctions}), this reads
\beq \label{mDK} m_{DK}^2 = { 1-3\omega + n -{ \kappa R_n \over n \rho} \left( {1-2n \over 1-4n}{R \over R_n}\right)^{1-n} \over 3 (n-1) } R = { 2 n (2 + n - 3 w) -1 \over 6n (n-1) } R > 0 ~~,\eeq
after using Eq. (\ref{rhoandomega}). Thus, for a negative exponent $n<0$ this requires that $n < - (2-3\omega + \sqrt{3(2-4\omega+3\omega^2})/2$: since visible matter is almost pressureless, the constraint $\omega \sim 0$ finally leads to $n \lesssim - (1 + \sqrt{3/2}) \approx -2.2$, which is clearly incompatible with the visible and dark matter profiles discussed in the following section.

As such, we arrive at the conclusion that $m_{DK}^2$ is negative for the physical range of values for the exponent $n$ and equation of state parameter $\omega$: a similar conclusion was drawn in Ref. \cite{DMOBJP}, which this study now generalizes through the inclusion of a non-vanishing visible matter pressure.

This is also similar to the result obtained in Ref. \cite{Faraoni:2007sn}, where the more general condition $F_1' + F_2' \cl > 0 $ is obtained --- which yields $n(1-n)>0$, clearly incompatible with a negative exponent $n$ (notice that our result is more convoluted since the use of a conformal transformation implies that $f_1(R) \neq R$, as depicted in the form (\ref{ffunctions})).

Following Ref. \cite{DMOBJP}, we now recall that the analytical treatment leading to the identification of a putative Dolgov-Kawasaki instability resorts to the expansion of the equations of motion in a cosmological setting, thus considering a background, time-evolving scalar curvature $R=R_0(t)$ where small spatial perturbations may arise.

In the present context, the scalar curvature is not homogeneous, but clearly varying with the distance to the center $r$ --- so that greater care should be taken when applying the aforementioned mechanism. Indeed, for the usual treatment to apply, we expect that the curvature varies softly over the characteristic lengthscale of the instabilities $m_{DK}^{-1}$: this can be translated into the condition
\beq \left|{R(r) \over R'(r) }\right| \gg \left|m_{DK}\right|^{-1} ~~.\eeq 

Disregarding factors of order unity, this reads $ R^3(r) \gg [R'(r)]^2$: since Eq. (\ref{dmvariables1}) tells us that the scalar curvature scales as the mimicked dark matter density, $\ka R(r) \sim \rho_{dm} \sim \rho_{dm(f)} (r/r_f)^{1/2n}$, this becomes 
\beq \left({r \over r_f}\right)^{2+{1\over 2n}} \gg {\ka \over \rho_{dm(f)} r_f^2}~~,\eeq
where $\rho_{dm(f)} = \rho_{dm}(r_f)$ is the density of the mimicked dark matter at the edge of the galaxy. Using typical orders of magnitude $\rho_{dm(f)} \sim 10^{-22}$ kg/m$^3$ and $r_f \sim 10-100$ kpc finally yields the constraint $\left({r \over r_f}\right)^{2+{1\over 2n}} \gg 10^7$. Since $r<r_f$, this requires that the exponent on the r.h.s. of the above expression is large and negative, so that the NMC exponent must be extremely small, $n\lesssim 0$.

As all visible and mimicked dark matter profiles considered in the following section require an NMC exponent that does not fulfil the above (namely $n=-1/7$, $-1/6$ or $-1/4$), we naturally obtain the same conclusion as found in Ref. \cite{DMOBJP}: that the overall procedure leading to the assessment of Dolgov-Kawasaki instabilities does not apply in the present scenario.

A more complete study considering the boundary matching with an evolving cosmological background should shed further light into the stability of the solutions obtained. Speculatively, we anticipate that the relevant timescale for the evolution of perturbations should be related to the Hubble time and, as such, allow for at least semi-stable solutions during the relevant timespan.

\subsection{Standard DM Profiles}
\label{DMprofiles}

In order to compare the obtained solution with well known dark matter profiles, namely the Navarro-Frenk-White (NFW) \cite{1996ApJ...462..563N} and isothermal \cite{Suto:1998xs} distributions, we approximate the solution \eqref{redcurv} with
\beq y(x) \sim y_f \left({2	x \over x_f}\right)^{1\over 2n}~~, \label{aprox3} \eeq
which is valid suficiently inside the galaxy, $x \ll x_f$. Thus, we obtain a direct translation between the slope of the dark matter distribution $\rho_{dm} \sim y$ and the NMC exponent $n$.

\subsubsection{Navarro-Frenk-White dark matter profile:}

In the outer galactic regions, the NFW dark matter density profile scales as $ \rho_{dm}(r) \propto r^{-3} $. Thus,  Eq. \eqref{aprox3} immediately yelds
\beq {1\over 2n} = -3 \to n= - {1\over 6}~~, \eeq 
so that visible matter scales as
\beq \varrho \propto y^{1-n} \propto x^{-{7/2}}~~. \eeq 
\subsubsection{Isothermal dark matter profile:}

By the same token, the isothermal profile falls with $ \rho_{dm}(r) \propto r^{-2}$, so that
\beq {1 \over 2n} = -2 \to n = -{1\over 4}~~, \eeq
and visible matter behaves as
\beq \varrho \propto x^{-{5/2}}~~. \eeq

Comparing with the Hernquist profile \cite{1990ApJ...356..359H} for luminous matter, where $ \varrho \propto x^{-4} $, we conclude that the NFW profile appears to provide the closest match.
Conversely, we compute that the NMC exponent $ n $ which yields the exact Hernquist profile is given by $ n = -1/7$.

\subsection{Mass} \label{masss}

When computing the mass of our spherical object, one needs to be careful with the limits of integration: indeed, since the validity of solution \eqref{redcurv} is verified only in the galactic halo where dark matter dominates, one cannot simply integrate from the center to a given radius $ r $.

Hence, we define an inner radius $ a $ marking the region where dark matter starts to dominate over visible matter: since we do not have a description of the interior region of the galaxy, where visible matter is dominant and the relaxed regime \eqref{re0} is no longer valid, this crossover radius cannot be extracted from the model: instead, it should stem from a more realistic simulation of the overall dynamics, which falls outside the professed scope of this work.

Using definition \eqref{metrics}, the mass component $M_i$ enclosed in the halo region $a < r < r_f$ is given by
\beq M_i = 4 \pi \int_{a}^{r_f}\tilde{\rho}_i(r) r^2 dr~~.\eeq 
Resorting to the Eq. \eqref{aprox3}, we get
\begin{eqnarray} \rho(r) &\sim& {1 \over 1-4n} 2 \kappa R_2 \left(R_f \over R_2\right)^{1-n}\left(2r \over r_f\right)^{1-n \over 2n} \nonumber~~, \\ \rho_{dm}(r) &\sim& {1-n \over 1-4n} 2\kappa R_f \left(2r \over r_f\right)^{1\over 2n}~~, 
\end{eqnarray}
and the contribution of dark matter to the mass is
\beq M_{dm} = h_n {\tilde{R}_f r^3_f c^2\over G}\left[1-\left(a \over r_f\right)^{1+6n \over 2n}\right]~~,	 \label{totaldmmass}\eeq
for $ n \neq -1/6 $, where
\beq h_n \equiv {2^{1 \over 2n}n (1-n) \over (1-4n)(1+6n)}~~.\eeq
Notice that the condition $n < -1/6$ is required to keep the mass positive defined.

For $ n = - 1/6 $, it is easy to see that the dark matter mass has a logarithmic dependence (a characteristic of the NFW profile),

\beq M_{dm} = {7 \over 160} \tilde{R}_f r_f^2 { r_f c^2 \over G} \ln \left( {r_f \over a} \right) \label{dmmass2}~~.\eeq

This can also be obtained by taking the limit $n\to -1/6$ of Eq. (\ref{totaldmmass}).

For any value of the exponent $n \neq -1/5$ (which does not correspond to any profile of interest and is thus disregarded), the visible mass enclosed within the mimicked dark matter halo is given by
\beq M_v =  h_n'\left(\tilde{R}_f \over R_n\right)^{-n} {\tilde{R}_f r_f^3 c^2 \over G}\left[1-\left({a \over r_f}\right)^{1+5n \over 2n}\right]~~, \label{totalvismass}\eeq
\noindent with
\beq h'_n \equiv {2^{1-n\over 2n}n \over (1-4n)(1+5n)}~~. \eeq
For $n=-1/5$ we see that $h'_n$ diverges, although the visible mass is given by
\beq M_v = {5 \over 144} \left(\tilde{R}_f \over R_{-1/5}\right)^{1/5} {\tilde{R}_f r_f^3 c^2 \over G} \log \left({ r_f \over a}\right)~~. \label{totalvismass2}\eeq
As this value for the exponent $n$ does not correspond to any relevant visible or dark matter distribution, we shall not consider it further.

The ratio between dark matter and visible mass is thus

\beq {M_{dm} \over M_v} = {h_n \over h_n'}\left(\tilde{R}_f \over R_n\right)^n {1-\left(a\over r_f\right)^{1+6n \over 2n} \over 1-\left(a\over r_f\right)^{1+5n \over 2n}}~~, \label{massratio1} \eeq
for $ n \neq -1/6 $; since $h'_n \sim h_n$ and $a \lesssim r_f$, the above is large if  $ (\tilde{R}_f / R_n)^n = y_f \sim f_2 \gg 1$: a dominance of dark over visible matter implies a strong NMC, as desired.

Finally, for $n = -1/6$, we get

\beq {M_{dm} \over M_v} = {7 \over 6 \sqrt{2}} \left( {R_{-1/6} \over \tilde{R}_f} \right)^{1/6} {\ln\left(r_f \over a\right) \over \sqrt{r_f \over a} -1 }~~. \label{massratio2} \eeq

\section{Model Constraints} \label{validity}

In this section, we aim to compare the expressions found in the preceding section with observed values. Since our solutions were found by the assumption that our object has spherical symmetry, we adopt the dataset reported in Refs. \cite{Gerhard:2000ck,Kronawitter} for galaxies of type E0, which display low eccentricity. The relevant distances $(a, r_f)$ and enclosed visible and dark matter masses $M_v$ and $M_{dm}$ were inferred from the corresponding rotation curves and are depicted in Table \ref{tab:table1}.

\begin{table}
	\centering
	\begin{tabular}{c|c|c|c|c|c}
		NGC  & $  M_v(10^{10}M_\odot) $ & $M_{dm}(10^{11}M_\odot)$ & $ r_f(kpc) $ & $ a(kpc)$ & $y_f (10^{-6})$  \\[2pt] \hline
		2434 &           4.6            &     1.9     &      18      &   5.1 & 7    \\
		5846 &           36.4           &     8.5     &      70      &   21.1 & 221  \\
		6703 &           5.6            &      1.2    &      18      &   2.7   & 697 \\
		7145 &           3.8            &       1.4   &      25      &   4.0   & 21 \\
		7192 &           7.1            &      2.7    &      30      &   4.1   & 18 \\
		7507 &           7.9            &       3.5   &      18      &   4.7   &  5\\
		7626 &           30.7           &      8.7    &      50      &   11.0  &  99
	\end{tabular}%
	\caption{Table of visible $M_v$ and dark matter $M_{dm}$ masses, dark matter halo radius $r_f$, crossover distance $a$ and reduced curvature end value $y_f$ obtained by performing a fit of the rotational curves of selected galaxies with  the best fit exponent $n=-1/6$.}
	\label{tab:table1}
\end{table}

For that purpose, we eliminate $ \tilde{R}_f $ in Eqs. (\ref{massratio1}) and (\ref{massratio2}) by resorting to  \eqref{totaldmmass} and \eqref{dmmass2} yielding, respectively

\beq \label{masscomparing} {M_{dm} \over M_v} = {  h_n^{1-n} \over h'_n} \left( { r_f c^2 \over GM_{dm} } r_f^2 R_n \right)^{-n}  \left[ 1 - \left({a \over r_f}\right)^{1+5n\over 2n}\right]^{-1} \left[ 1 - \left({a \over r_f}\right)^{1+6n\over 2n}\right]^{1-n}
~~, \eeq
for $n\neq -1/6$, and
\beq  {M_{dm} \over  M_v} = {7 \over 12} \left( {7 \over 20} {  r_f c^2 \over  GM_{dm} } r^2_fR_{-1/6} \right)^{1/6} { \left[ \ln \left({ r_f \over a} \right)  \right]^{7/6}  \over \sqrt{ r_f \over a} -1 } ~~, \label{masscomparing2}\eeq
for $n=-1/6$. As before, this expression may be obtained as the limiting case of Eq. (\ref{masscomparing}) when $n \to -1/6$, so the transition to a logarithmic dependence is smooth.

Notice that, in this study, the two lengthscales $a$ and $r_f$ are not derived from the model: $r_f$ marks the endpoint of each galaxy's rotation curve, while $a$ signals the onset of the relaxed regime $F_1+F_2\cl = {\rm const.}$, where we find that the effect of the NMC can be interpreted as a dominant dark matter component; for interior regions $ r < a$, visible matter dominates and the relaxed regime should not valid.

Naturally, a further, overarching study should consider all distances and numerically solve the unconstrained equations of motion (\ref{fieldequation}): as mentioned after Eq. (\ref{curvaturetilde}), it was previously found that the relaxed condition (\ref{re0}) indeed arises dynamically and is valid for large distances $r >a$ \cite{DMOBJP}; however, it is still unclear how the crossover distance $a$ is in principle related to the model parameters $n$ and $R_n$. 

With the above in mind, we use the above expressions (\ref{masscomparing}) and (\ref{masscomparing2}) for the dark to visible mass ratio to compute the latter for each of the considered galaxies, using the respective values for the lengths $a$ and $r_f$ obtained from Refs. \cite{Gerhard:2000ck,Kronawitter}.

We then sweep over the allowed range for the exponent $n<0$ and characteristic curvature scale $R_n$ and iteratively compute the obtained expressions for $M_{dm}/M_v (n,R_n)$ for each galaxy. By comparing with the observed value extracted from Refs. \cite{Gerhard:2000ck,Kronawitter}, the correlation coefficient $r^2$ is then computed for each pair $(n,R_n)$ and the best fit scenario identified as the specific pair $(n^*,R_n^*)$ that maximizes the former.

We find that the value for the adjusted correlation coefficient $r^2$ increases continuously as one arbitrarily approaches $n =-1/6$ from both sides, for any given value of $R_n$: the best fit is then found when $n=-1/6$ exactly.

The correlation coefficient for this best fit scenario is $r^2 = 0.86$: this relatively low value can reflect {\it e.g.} small deviations from sphericity in the selected type E0 galaxies, as well as unmodeled localized features and inhomogeneities.

The best fit for the characteristic curvature scale was found to be $R_n \approx 5 $ Mpc$^{-2}$, although with a large uncertainty stemming from the relative lack of sensibility of \eqref{masscomparing2} to its value: as such, we conservatively assess only that $R_n \sim $ Mpc$^{-2}$. 

Finally, it is also interesting to verify that the assumption of a strong NMC holds. For that, we resort to Eq. \eqref{massratio2} ({\it i.e.} the best fit scenario $n=-1/6$) and solve it for $ \tilde{R}_f /R_{-1/6} $. This yields 

\beq y_f = {\tilde{R}_f \over R_{-1/6}} = {1 \over 8}\left( {7\over 6}{M_v \over M_{dm}}{\ln\left(r_f \over a\right) \over \sqrt{r_f \over a} - 1} \right)^6 \label{yf2}~~, 
\eeq
and the values for distinct galaxies can be found in the last column of Table \ref{tab:table1}: the results confirm that the assumed perturbative regime $y \sim y_f \ll 1$ is indeed valid.

\section{Conclusions} \label{conc}

In this work, we showed that the rich phenomenology of a model endowed with a nonminimal coupling between curvature and matter allows for the possibility of mimicking dark matter profiles for galaxies which exhibit spheric symmetry. This result was obtained by assuming a relaxed regime, interpreted as a fixed point condition, which allowed us to relate the Lagrangian density with the curvature, replacing the need for an additional EOS.

By assuming a power law form to the NMC function $ f_2(R) $ motivated by the previous work reported in Ref. \cite{DMOBJP}, we have solved the relevant equation in the perturbative regime and characterized the ensuing solutions: in particular, we confirm the relation between visible and dark matter density profiles found in that study and further refine it by explicitly deriving their individual dependence on the exponent $n$ of the NMC function: we recall that Ref. \cite{DMOBJP} only provided the translation between visible and dark matter profiles, but did not account for why these adopt a particular radial dependence.

We also show that the NMC must be strong, $f_2 \gg 1$ if the mimicked dark matter is to dominate at large distances: this translates into a perturbative regime where the scalar curvature is much smaller than the characteristic curvature scale $R \ll R_n$. Furthermore, we have shown that this condition does not violate the appropriate energy conditions.

Finally, the enclosed masses were obtained and used to compare with observation, showing that the NFW profile for the mimicked dark matter component is favored, given by an exponent $n = -1/6$ (or an arbitrarily close value). The characteristic curvature scale was found to be of the order of $R_n \sim 1$ Mpc$^{-2}$, although the quality of the fit is not sufficient to fix it more accurately.

Given the interesting results obtained, it is tempting to assume that all dark matter stems from the dynamical effect of the NMC between curvature and matter. However, the striking evidence of the separation between dark and visible matter provided by the Bullet cluster further complicates the issue \cite{bullet}.

Indeed, if dark matter is not a real matter type, but merely reflects the enhanced gravity of visible matter due to the model under scrutiny (or, indeed, any extension of GR), then it is plausible that it should inherit the spatial symmetry of visible matter: in other words, if visible matter interacts and produces the shock wave directly observed in the Bullet cluster, the ensuing gravitational profile should also exhibit this feature, instead of maintaining an approximate spherical symmetry. This, however, could be tackled by assuming that the NMC is more strongly coupled to neutrinos, which do not interact, than to visible matter: such possibility should be further investigated.
	
	\ack The authors thank the referees for their useful remarks and criticisms.

\end{document}